# Fossil Turbulence Revisited

Carl H. Gibson

*Departments of Applied Mechanics and Engineering Sciences
and Scripps Institution of Oceanography
University of California at San Diego
La Jolla, CA 92093-0411*

## Abstract

A theory of fossil turbulence presented in the 11th Liege Colloquium on Marine turbulence is "revisited" in the 29th Liege Colloquium "Marine Turbulence Revisited". The Gibson (1980) theory applied universal similarity theories of turbulence and turbulent mixing to the vertical evolution of an isolated patch of turbulence in a stratified fluid as it is constrained and fossilized by buoyancy forces. Towed oceanic microstructure measurements of Schedvin (1979) confirmed the predicted universal constants. Universal constants, spectra, hydrodynamic phase diagrams (HPDs) and other predictions of the theory have been reconfirmed by a wide variety of field and laboratory observations. $A_T$ (turbulent activity coefficient) versus C (Cox number) HPDs classify microstructure patches as active, active-fossil, and fossil turbulence. So do Froude-Reynolds number $Fr/Fr_o$ versus $Re/Re_F$ plots. Both HPDs show most oceanic microstructure patches are fossilized. The oceanic microstructure community has not yet adopted the fossil turbulence paradigm, for reasons that include a variety of misconceptions about stratified turbulence and turbulent mixing. Confusion of fossilized microstructure with turbulent microstructure leads to vast underestimates of average dissipation rates of kinetic energy and scalar variance , and therefore vast underestimates of vertical fluxes in most ocean layers. Fossil turbulence theory has many applications; for example, in marine biology, laboratory and field measurements suggest phytoplankton species with different swimming abilities adjust their growth strategies differently by pattern recognition of several days of turbulence-fossil-turbulence dissipation and persistence times above threshold values, signaling a developing surface layer sea change. In cosmology, self-gravitational structure masses are interpreted as fossils of primordial hydrodynamic states.

## 1. Introduction

The fossil turbulence phenomenon, comprising various persistent "footprints" of previous turbulence events, has been recognized independently in several fields, including fluid mechanics, powder metallurgy, cosmology, meteorology, and oceanography. The term "fossil turbulence" was perhaps first mentioned in print by George Gamov (1954) who suggested that the space and mass distributions of galaxies were fossils of powerful primordial turbulence driven by the Big Bang because density fluctuations of the turbulence would influence the formation of such gravitational structures. Ozernoy (1978) summarizes attempts to relate Kolmogorov (1941) universal similarity theories of turbulence to various astrophysical structures. However, this effort was abandoned when measurements from the COsmic Background Explorer (COBE) satellite revealed a very small level of temperature T, and therefore velocity v, fluctuations in the microwave background radiation field ( v/c  T/T  $10^{-5}$), where c is light speed, showing that turbulence of the primordial gases emerging from the plasma-gas transition 300,000 years after the big bang must have been extremely weak. Fossil-turbulence-nonturbulence structures were indeed formed at this time, but at earth-moon mass-scales $10^{19}$ smaller than those of galaxies, Gibson (1996, 1997). Proto-galaxies and proto-galaxy-clusters were formed earlier in the superviscous plasma epoch by this nonlinear, non-acoustic, gravitational condensation theory.

Oceanographic and atmospheric fossil turbulence was first reported at a fossil turbulence workshop organized by John Woods (1969). In the atmosphere, fossil turbulence manifests itself by radar returns from refractive index microstructure (termed "angels") persisting far downstream of mountains, and becoming vertically anisotropic due to stratification and therefore clearly nonturbulent, Richter (1969). Jet aircraft contrails in stably stratified layers often persist for hours compared to only minutes required to damp their largest scale turbulence. Skywriting is another commonly observed example of stratified atmospheric fossil turbulence. Skywriting won't work below the density inversion layer of the atmosphere because the smoke disperses too rapidly. Stewart (1969) reports from submarine measurements of simultaneous velocity and temperature gradients that most temperature microstructure patches observed must be fossil turbulence because few were accompanied by velocity microstructure vigorous enough to produce them. Woods (1969) observed that dye microstructure formed from internal wave turbulence in dyed layers persists indefinitely without vertical collapse, long after all turbulence overturning motions have ceased. Active (overturning) turbulence grows by entrainment until buoyancy forces become dominant. The remnants are stratified fossil turbulence or, in these cases, fossil-refractive-index-turbulence, fossil-temperature-turbulence, and fossil-dye-turbulence.

Panton (1984, p. 392) describes viscous fossil-smoke-turbulence, formed by a cylinder wake leaving smoke patterns in a wind tunnel that persist after viscous damping of the turbulence. When a smoke wire is placed near the cylinder the scrambled smoke patterns persist far downstream where no eddies exist, as demonstrated by moving the smoke wire far downstream. Viscous fossil turbulence occurs when the turbulence is damped by viscous forces at small scales, leaving the weakly diffusive smoke tracer of previous eddies as a fossil.

Stratified fossil turbulence is formed when buoyancy forces damp the largest eddies of a turbulence patch. Anyone can observe stratified fossil turbulence formation by rapidly pouring cold milk into hot tea or coffee. Turbulence grows from small scales to large as the turbulent milk jet entrains the hot fluid and then fossilizes to a bobbing, fossil-milk-turbulence patch that does not collapse and completes the milk mixing initiated by the turbulence. Nasmyth (1970) demonstrated the qualitative evidence of fossil turbulence noticed by Stewart (1969), but unfortunately used a frozen-fossil definition of fossil turbulence whose existence is physically impossible in the continuously moving ocean interior. A more realistic definition of fossil turbulence is discussed in Section 3. Surprisingly, little mention exists in the literature about fossil turbulence during the decade 1970-1980 when microstructure measurements were first carried out extensively in the world's oceans, other than Schedvin (1979) who demonstrated the phenomenon from vertically profiling towed body measurements of temperature microstructure to the diffusive scale.

Schedvin's data are included in the Gibson (1980) fossil turbulence theory as its experimental basis.

In some cases fossil turbulence has not been properly recognized or studied by oceanographers because their published definitions of turbulence are so broad that the buoyancy dominated motions of fossil-vorticity-turbulence and even the scalar fluctuations of fossil-temperature-turbulence or fossil-salinity-turbulence are classified as turbulence. Another problem is the common misconception, prevalent not only among oceanographers, that turbulence always cascades from large scales to small. Horizontally polarized "two-dimensional" turbulence in the ocean and atmosphere, which obviously grows from small scales to large, is said to undergo a "reverse cascade". However, it should be equally obvious that three dimensional turbulence also cascades from small scales to large, and for the same reasons. All turbulence is driven by inertial vortex forces and eddy pairings and most of the vorticity forms at the smallest scale, even in stratified flows. The recommended definition of turbulence is given in Table 1 and Section 2.1.

The signature of stratified fossil turbulence is a density microstructure patch with large overturning scales and small ; more precisely, with maximum Thorpe overturning scale $L_T$ greater than 60% of the Ozmidov scale $L_R = (\ /N^3)^{1/2}$, where $L_T$ is the maximum vertical displacement of density from a monotonic profile in the patch,   is the viscous dissipation rate, N is the stratification frequency $[g(\ /\ z)/\ ]^{1/2}$, g is gravity,   is density, z is down, and the density gradient ( / z) is averaged over a vertical scale larger than the patch. Laboratory measurements confirm that the fossilization criterion is $L_T$   $0.6L_R$, Gibson (1991d). If turbulence actually ever did cascade from large scales to small following the cascade myth, then a large microstructure patch with small   could be either fossil turbulence or a large turbulent eddy at such an early stage that the small scale structure has not yet developed. Gregg (1987) makes this claim based on tilt-tube evidence of Thorpe (1968) that gives the illusion that the large Kelvin-Helmholtz (KH) billows formed are nonturbulent. Gregg suggests that the first turbulence in such billows forms at a late stage due to gravitational collapse, and that after this the turbulence and the microstructure collapse together and vanish without a trace. Thus, no fossil turbulence forms in this model.

The Gregg (1987) model of ocean turbulence and fossil turbulence is representative of that used by many physical oceanographers at present, and is incorrect and misleading as discussed by Gibson (1988b). The jelly roll structure of the Thorpe (1968) billows is an optical illusion of non-turbulence, similar to the initial laminar morphology of wing-tip vortices produced by airplane wings rolling up turbulent boundary layers, and followed by a similar violent "vortex breakdown". Lack of radial mixing in both structures is caused by powerful Coriolis forces of the roll-up that initially inhibit radial turbulence because the Hopfinger scale $L_H = (\ /\ ^3)^{1/2}$ is smaller than the thin concentric circular layers of the billows. As the vortex angular velocity   decreases with increasing due to vortex line stretching, $L_H$ becomes larger than the billow size so that the radial turbulence of the rolled up boundary layers is released and rapidly mixes the billow (the vortex breakdown mechanism) in the radial direction, De Silva et al. (1996). Gravity has nothing to do with the formation of turbulence in such billows. KH billows do not collapse due to buoyancy except for small Reynolds number patches, or else slightly at very late stages of fossilization for initially high Reynolds number patches. They continue to grow slowly even after buoyancy forces cause fossilization at the largest eddy scales.

Based on comparisons of published oceanic microstructure data to fossil turbulence theory using hydrodynamic phase diagrams (HPDs) discussed in Sections 3 and 4 below, a partial list of references where physical oceanographers have confused fossilized turbulence with active turbulence is: Osborn and Cox 1972; Osborn 1980; Caldwell et al. 1980; Oakey 1982; Gregg 1976, 1977, 1980, 1987, 1989; Dillon 1984; Gargett et al. 1979, Gargett et al. 1981, Gargett 1985, 1989; Hebert et al. 1992; Peters, Gregg and Sanford 1994; Wijesekera and Dillon 1991, 1997; Toole, Polzin and Schmitt 1994; Kunze and Sanford 1996; Polzin, Toole, Ledwell and Schmitt 1995, 1997. The confusion derives from a variety of misconceptions about turbulence, stratified turbulence, and fossil turbulence, which are compared to the facts according to Gibson 1980, 1981, 1982abc, 1986, 1987, 1988ab, 1990, 1991abcd, 1996, 1998 in Table 1.

**Table 1.** Misconceptions and Facts About Oceanic Turbulence and Fossil Turbulence

| Misconception | Fact |
|---|---|
| **1.** Turbulence always starts at large scales and cascades to small; for example, by the gravitational collapse of KH billows which form with small and no embedded small-scale turbulence. | Turbulence always starts at small scales and cascades to large. Consider jets, wakes, boundary layers and mixing layers. Tilt tube KH billows and wing tip vortices are optical illusions of large scale turbulent eddies that have formed prior to the formation of small scale eddies. Both of these structures contain rolled up turbulent boundary layers temporarily inhibited in the radial directions of the rolls by Coriolis forces. They don't collapse. |
| **2.** Anything that wiggles is turbulence. | Turbulence is an eddy-like state of fluid motion where the inertial-vortex forces of the eddies are larger than any other forces that tend to damp out the eddies. Microstructure patches can be unambiguously classified using hydrodynamic phase diagrams according to their hydrodynamic states as active, active-fossil, or completely fossil turbulence. |
| **3.** All oceanic turbulence is caused by breaking internal waves. | Dominant turbulence events in most layers are more likely to be caused by shear layers formed on fronts within 2D fossil turbulence structures resulting from inhomogeneous horizontal forcing. These turbulence events would therefore cause most small scale oceanic internal waves when buoyancy forces damp the turbulence events, rather than the reverse. |
| **4.** Waves don't mix. | Most of the mixing in the ocean occurs in fossilized turbulence microstructure patches where the energy of the mixing is supplied by fossil vorticity turbulence, which is a form of internal wave. Thus oceanic internal waves do most of the mixing, completing the microstructure processes started by turbulent stirring. |
| **5.** Fossil turbulence is not important in the ocean, and is not observed. | Fossil turbulence is crucially important to mixing and diffusion of the ocean in all layers. It is usually all that is observed for the largest patches of the layers, which are those most likely to dominate these processes. It provides a valuable parity check on undersampling dissipation rates due to intermittency. |
| **6.** Turbulence is not important to vertical diffusion in the ocean interior. | Turbulence almost certainly dominates vertical diffusion in all important layers of the ocean interior, surface, bottom, and edges. Evidence that it does not is likely to be the result of its vast undersampling. |
| **7.** Turbulence is homogeneous and stationary in layers over large horizontal distances for long times, so it is adequate to sample once vertically to characterize the turbulence in the layer. | Turbulence is extremely intermittent in space and time in horizontal layers, so it is necessary to sample horizontally and vertically within a given layer on scales larger than the largest horizontal or vertical eddy that ever occurs in the layer over time periods longer than the largest interval between layer storms. The most powerful events dominate the layer dissipation averages, and occupy a small fraction of space-time. |
| **8.** Turbulence and turbulent mixing in the ocean are not subject to universal similarity theories. | All evidence of departures from turbulence universal similarity theories can be attributed to misconceptions 1-7. The ocean is part of the universe, and is therefore subject to universal similarity laws of turbulence and turbulent mixing. |
| **9.** The maximum age of a microstructure patch is a small fraction of the stratification time $N^{-1}$. | Microstructure produced by turbulence persists in numerous hydrophysical fields much longer than the damping time $N^{-1}$ for active turbulence in a stratified fluid, depending mostly on the diffusivity of the field and the Reynolds number of the event. |

Consequences of the misconceptions in Table 1 are devastating to the reliability of oceanic turbulence studies and the credibility of their results and conclusions. By the first two misconceptions, the signature of fossil turbulence is not unique so that all microstructure is lumped together as turbulence and data sets containing many patches are mistakenly taken as representative of the turbulence process even though few or none of the patches may actually be turbulent at the time of sampling. Fossil turbulence effects must be taken into account in the analysis of microstructure data from any stratified fluid with turbulence. This is not difficult, but is rarely done. Microstructure measurements should particularly be examined for the hydrodynamic state of those patches that dominate estimates of the average dissipation rates for the oceanic layer. If such patches are strongly fossilized, as expected since this has been the case for nearly all dropsonde data sets so far reported (See Figure 6 below), then the turbulence processes of the layer are likely to be vastly undersampled and the average dissipation rates and vertical fluxes vastly underestimated.

The present paper will first discuss turbulence theory in Section 2, since much of the confusion and most of the misconceptions of physical oceanographers about fossil turbulence can be traced to flaws in the definitions and cascade direction of turbulence proposed in commonly accepted turbulence descriptions in the fluid mechanics literature, usually developed with homogeneous, isotropic laboratory flows in mind rather than the stratified, rotating fluids of the ocean and atmosphere. Section 3 will define fossil turbulence and show how it can be identified using hydrodynamic phase diagrams. Section 4 will discuss fossil turbulence measured in the laboratory and ocean using a summary HPD, and will give two major oceanic applications, the resolution of the dark mixing paradox and a proposed working hypothesis for the response of phytoplankton to turbulence and turbulence intermittency in surface layers of the sea. Section 5 shows the close analogy between stratified turbulence in the ocean and in the self-gravitational systems of astrophysics and cosmology. Finally, a summary and conclusions is provided in Section 6.

## 2. Turbulence theory

In this section we address the first two misconceptions of Table 1; that is, the idea that any microstructure in velocity, temperature, or other hydrophysical field is sufficient evidence that the state of fluid motion is turbulence ("anything that wiggles is turbulence") and the idea that turbulence is formed first at large scales and then cascades to small scales. Fossil turbulence could not be distinguished from turbulence if either of these ideas were true.

### 2.1 Definition of turbulence

Turbulence is a property of fluid flows that has been notoriously difficult to define. Libby (1996, p2) uses a wide spectrum of velocity fluctuations to distinguish turbulence from "unsteady laminar flow". Frisch (1995) simply emphasizes the requirement of high Reynolds number. Syndromic definitions have been offered (Stewart 1969; Tennekes and Lumley 1972) that list various commonly accepted properties of turbulence. Tennekes and Lumley list irregularity, diffusivity, large Reynolds number, three-dimensional vorticity fluctuations, and large dissipation rates as distinguishing properties of turbulence. Individually the properties listed may be neither necessary nor sufficient. Only when taken together do they add up to a convincing diagnosis that turbulence exists, just as a disease is diagnosed by matching symptoms to the syndrome or criminal culpability is established by the preponderance of the evidence.

Unfortunately, such syndromic definitions or non-definitions give the impression that turbulence has no intrinsic scientific basis and may be defined according to the whim of the user or the needs of the situation. Like pornography, "you will know it when you see it." A more precise definition is required for the complex flows of natural fluids based on the crucial non-linear term, the inertial-vortex force of the momentum conservation equations that causes the turbulence in the first place and sustains the cascade to large scales. This turbulence definition is crucial for oceanic conditions where buoyancy and Coriolis forces at the largest scales rapidly convert rare active turbulence patches to fossil turbulence remnants, Gibson (1980, 1990, 1991abcd). Viscously damped turbulence leading to viscous fossil turbulence of large Schmidt number tracers like smoke and dye is a relatively unimportant phenomenon in non-stratified, non-rotating turbulence fluid mechanics, appearing perhaps as a footnote to flow visualization studies. However, fossil turbulence dominates the stratified, rotating turbulence and turbulent mixing of the ocean, atmosphere, and all other natural fluids of the cosmos. Fossil turbulence dynamics represent the nonlinear coupling mechanism between turbulence and small scale internal gravity waves, as well as many other wave motions from Coriolis, surface tension, and other forces.

> **Definition:** *Turbulence is defined as an eddy-like state of fluid motion where the inertial-vortex forces of the eddies are larger than any of the other forces which tend to damp them out.*

Note that irrotational flows, with zero vorticity, are non-turbulent by definition because the inertial-vortex force per unit mass $\mathbf{v} \times \boldsymbol{\omega}$ is zero, where $\mathbf{v}$ is the velocity and $\boldsymbol{\omega}$ is the vorticity. The cascade of energy from large scales to small that supplies the energy of turbulence is induced by turbulence, but the flow itself is non-turbulent, ideal, inviscid, and irrotational. The turbulence eddy-like motions cascade from small scales to large with feed-back, extracting kinetic energy from the entrained non-turbulent surrounding fluid at all scales. Flows like internal waves that are dominated by buoyancy forces are non-turbulent by definition, even though the waves are rotational, non-linear, random, and may even be caused by turbulence. Flows may be eddy-like, non-linear, and random, like Coriolis-inertial waves formed when two dimensional turbulence is damped by Coriolis forces, but such waves are not turbulent by definition because the Coriolis forces dominate the inertial-vortex forces. Eddy-like flows in cavities of boundary layers (see plates 10 and 14 of Van Dyke 1982) are random, rotational, and diffusive, but are not turbulence by definition because they are dominated by viscous forces.

The term "turbulence" as defined in oceanography has evolved to become progressively narrower over time. Fifty-five years ago, Sverdrup et al. (1942) asserted that most of the ocean was turbulent, but today most oceanographers believe that at any instant of time only a small fraction of the ocean is fully turbulent: about 5% based on the volume fraction of temperature microstructure typically encountered by microstructure sensors in oceanic interior layers (Washburn and Gibson 1982, 1984). The reason is mostly due to a shift in the accepted definition of turbulence among most oceanographers. However, most microstructure patches are fossilized, so that the actual turbulence fraction is much less than 5% based on the classification of measured oceanic microstructure patches as "actively turbulent", "active-fossil", or "fossil turbulence" using hydrodynamic phase diagrams (discussed below). The oceanic fossil-turbulence to active-turbulence fraction approximately equals the Reynolds number ratio $Re_0/Re_F$ of the dominant turbulence events of a particular layer, which varies from about $10^3$ in surface layers to $10^4$ or more in abyssal layers, with a maximum of about $10^7$ near the equator.

## 2.2 Turbulence and fluid forces

The behavior of turbulence in the stratified, rotating ocean is best illustrated by the conservation of momentum equations written in the form

$$\frac{\partial \mathbf{v}}{\partial t} = \mathbf{v} \times \boldsymbol{\omega} + \mathbf{v} \times 2\boldsymbol{\Omega} - \nabla B + \nabla \cdot (\boldsymbol{\tau}/\rho) + \mathbf{b} + \ldots \qquad (1)$$

where $\mathbf{v}$ is the velocity, t is time, $\boldsymbol{\omega} =$ curl $\mathbf{v}$ is the vorticity, $\boldsymbol{\tau}$ is the viscous stress tensor, and the density $\rho$ is assumed constant. The Bernoulli group of mechanical energy terms $B = v^2/2 + p/\rho + gx_3$, where p is the pressure, g is gravity, and $x_3$ is up. B is constant for steady, irrotational, inviscid flows (the Bernoulli equation). The buoyancy force is $\mathbf{b}$. The Coriolis force is $\mathbf{v} \times 2\boldsymbol{\Omega}$, where $\boldsymbol{\Omega}$ is the angular velocity of the earth. Electromagnetic, surface tension, and other forces are omitted for simplicity. Turbulence occurs when the inertial-vortex forces $\mathbf{v} \times \boldsymbol{\omega}$ per unit mass exceed the viscous forces $\nabla \cdot (\boldsymbol{\tau}/\rho)$ per unit mass, if the other forces are negligible. The ratio of inertial-vortex forces to viscous forces is the Reynolds number $Re = UL/\nu$, where U is a characteristic velocity and L is a characteristic length scale of the flow. Re may also be considered the ratio of the eddy diffusivity UL to the molecular diffusivity, which shows why turbulent flows are strongly diffusive compared to viscous flows. Similar dimensionless groups arise from the other terms in (1). All of these must also be smaller than $\mathbf{v} \times$ for the flow to be turbulent. For example, the ratio of **b** to $\mathbf{v} \times$ is the Richardson number Ri $N^2/(\partial U/\partial z)^2$, where N is the intrinsic, or Väisäilä, frequency (the bobbing frequency of a displaced particle) of a stratified fluid, $N^2$ $g(\partial\rho/\partial z)/\rho$, $\rho$ is density, g is gravity, U is the velocity in the horizontal direction, and z is down. Ri must be less than some critical value for a flow to be turbulent, or equivalently its inverse the Froude number Fr = $1/(Ri)^{1/2}$ must be greater than a critical value. For rotating flows the Rossby number Ro = (inertial-vortex/Coriolis) forces = $\mathbf{v} \times \omega / \mathbf{v} \times 2\Omega$ must also be super-critical.

## 2.3 Formation of turbulence

Turbulence arises because of shear instability. Any flow that produces a shear layer is likely to produce turbulence. The reason is that perturbations of the shear layer induce inertial-vortex forces in the direction of the perturbation, so that the perturbation grows. Such positive perturbation feedback is what is meant by the term "instability". Shear layers are produced by flow around obstacles, jets into stagnant bodies of fluid, boundary layers and mixing layers. Figure 1 shows a shear layer (mixing layer) and the evolution of velocity perturbations to eddy-like motions because they induce inertial-vortex forces in the perturbation direction.

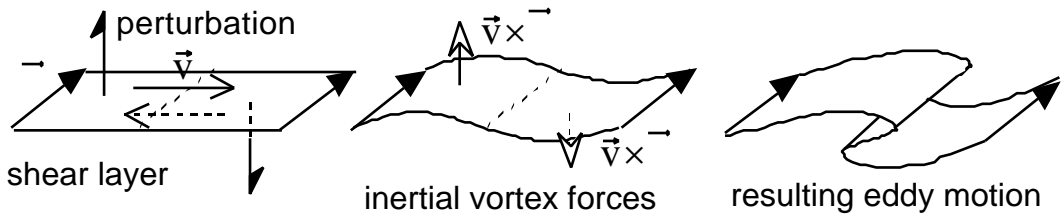

**Figure 1.** Fundamental mechanism of turbulence: perturbations on a shear layer cause inertial-vortex forces in the two opposite perturbation directions, with consequent eddy formation. Thin shear layers (vortex sheets) first thicken by viscous diffusion to the Kolmogorov scale before forming eddies; for example in laminar boundary layers.

Shear layers like that shown in Fig. 1 are unstable to perpendicular velocity perturbations U(L) at all scales L. However, the eddies that form first are those with the smallest overturn time T(L) = L/U(L), since usually U(L) is weakly dependent on L such as for turbulence with U(L) $\sim L^{1/3}$ so that T(L) $\sim L^{2/3}$ decreases with L from the Kolmogorov (1941) second universal similarity hypothesis. Viscosity limits the smallest eddy size to the Kolmogorov scale

$$L_K = (\nu^3/\varepsilon)^{1/4}, \qquad (2)$$

where $\nu$ is the kinematic viscosity of the fluid and $\varepsilon$ is the viscous dissipation rate

$$\varepsilon = 2\nu\, e_{ij}e_{ij}, \qquad (3)$$

$e_{ij}$ is the rate of strain tensor,

$$e_{ij} = (v_{i,j} + v_{j,i})/2 \qquad (4)$$

$v_i$ is the i component of velocity, and $v_{i,j}$ is the partial derivative of $v_i$ in the j direction, and repeated indices are summed over i and j = 1 to 3. It turns out that viscous forces prevent any eddies smaller than about $5L_K$. Such eddies have subcritical Reynolds numbers.

### 2.4 The turbulence cascade

Nearby eddies with the same average vorticity induce velocities such that they orbit about a point between them; that is, they pair to form a larger eddy. The transfer of energy and momentum between length scales (eddies) of comparable size is the turbulence cascade, and we see that on average it is from small scales to large, as shown in Figure 2 for an evolving shear layer, drawing energy from the irrotational fluid layers moving in opposite directions on opposite sides of the shear layer. Even if large eddy formation is strongly forced, as in wingtip vortices, the Kolmogorov eddies form first and grow to fill in the forced spectral gap in the universal Kolmogorov turbulence spectrum when Coriolis forces of the rotating structure permit (the process is termed vortex breakdown). The canonical poem of turbulence by L. F. Richardson; "Big whorls have smaller whorls that feed on their velocity, and smaller whorls have smaller whorls, and so on to viscosity (in the molecular sense)", is amusing but somewhat misleading. It should be replaced by doggerel with more fluid mechanical accuracy and a reference; such as, "Little whorls on vortex sheets, form and pair with more of, whorls that grow by vortex forces, Slava Kolmogorov!" ("Slava Kolmogorov!" means "Glory to Kolmogorov! in Russian).

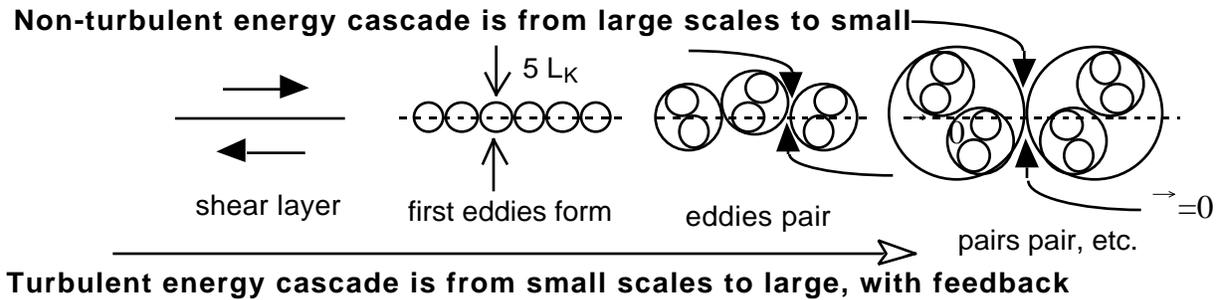

**Figure 2.** Schematic of the turbulence cascade process, from small scales to large. A non-turbulent energy cascade in the external, irrotational fluid from large scales to small is induced by the turbulence, and should not be confused with the turbulence energy cascade.

The external irrotational fluid entrained by a turbulent flow is the source of turbulence mass, momentum, and kinetic energy, and is also the source of the common misconception (Table 1) that turbulence cascades from large scales to small. The small to large cascade process illustrated in Fig. 2 would commonly be called an "inverse cascade", even though all turbulence flows develop according to the pattern shown, from small scales to large with intermediate scales in quasi-equilibrium. Note that any fluid properties such as temperature or nutrient concentrations are rapidly convected across the shear layer interface of Fig. 2 by the growing turbulence eddies. The maximum size of the turbulent eddies and the patterns of concentration fluctuations of entrained properties are limited by buoyancy forces in the vertical, Coriolis forces in the horizontal, or the extent of the shear layer, whichever constraint is reached first.

Figure 3 is a photograph of a high Mach number turbulent wake formed by a passing projectile. Note that the turbulence forms immediately at the smallest possible scales permitted by the viscosity of the fluid and cascades to larger scales as the eddies interact with each other.

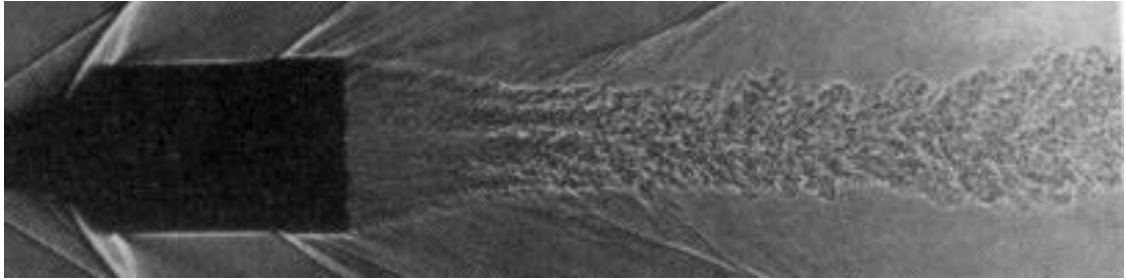

**Figure 3.** The wake of a high speed projectile shows small eddies of turbulence forming first, before the larger eddies downstream in the turbulent wake. In a stratified fluid the wake will grow to a vertical size about 0.6 $L_{R_o}$, where $L_{R_o}$ is the Ozmidov scale at the beginning of fossilization. The wake does not collapse (see Table 1), but gradually ceases vertical growth as it fossilizes at smaller and smaller scales due to buoyancy forces.

## 3. Fossil turbulence theory

A theory of fossil turbulence based on the evolution of an isolated turbulence patch was introduced by Gibson (1980). It was assumed that the turbulence of the patch starts at the smallest possible scale, the Kolmogorov scale $L_K = (\nu^3/\varepsilon)^{1/4}$, and grows by an eddy pairing process to the largest possible scale, the Ozmidov scale $L_{R_o} \equiv (\varepsilon/N^3)^{1/2}$ at the beginning of fossilization, where the dissipation rate $\varepsilon = \varepsilon_o$, $\nu$ is the kinematic viscosity and N is the Väisälä frequency of the ambient fluid. After beginning of fossilization the patch does not collapse as postulated by Gregg (1987), but the kinetic energy of the turbulence is converted to internal wave energy of trapped bobbing motions within the fossilized patch, with spectrum

$$\phi_u = 6.5 \, N^2 k^{-3} \tag{5}$$

interpreted as a universal saturated internal wave spectrum, or fossil vorticity turbulence spectrum, preserving the active turbulence kinetic energy as a fossil for wavenumbers $L_{R_o}^{-1} > k > L_R^{-1}$. The universal constant 6.5 was derived from universal similarity theory. The minimum turbulence dissipation rate when the turbulence reaches complete fossilization is

$$\varepsilon_F = 30 \, \nu N^2 \tag{6}$$

derived by Gibson (1980). Similarly, a saturated internal wave temperature spectrum

$$\phi_T = 0.7 \, (\partial T/\partial z)^2 \, k^{-3} \tag{7}$$

was derived, where T is the average temperature at depth z. Because the dissipation rates at beginning and end of the fossilization of the patch are very different but the kinetic energies are the same, the fossil persists much longer, by a factor of $\varepsilon_o/\varepsilon_F$. Since this ratio may be $10^4$ to $10^6$ for the dominant patches of interior oceanic layers, many thousands or millions of dominant patches should in principle be sampled before the data set may be considered representative. Only a small fraction of the fluid in such layers contains any microstructure, either active or fossil. Therefore, very long records of microstructure are required before reliable average dissipation rates can be estimated. If the dominant overturning patches for deep layers of the ocean interior are 10 meters in vertical extent and occupy a fraction $10^{-3}$ of the layer, it follows that about $10^9$ m of independent data record must be examined for the layer if $\varepsilon_o/\varepsilon_F$ is $10^5$ for a 50% chance of encountering an actively turbulent dominant patch. At a velocity of 1 m/s this would require about 30 years using a single sensor. Since the sensor would encounter a fossil patch about once every three hours, the motivation for developing a reliable system of hydropaleontology is clear. The best presently available data sets are typically smaller in length by factors of at least $10^4$; for example, Polzen et al. (1997).

Only the smallest scales of the fossil temperature turbulence spectrum are affected by the fossilization process, so these decay more rapidly as they are mixed by the internal wave motions of fossil vorticity turbulence. From the theory, this provides a means of estimating the beginning dissipation rate of the fossil $\varepsilon_o$

$$\varepsilon_o \approx 13 D C_x N^2 \tag{8}$$

where D is the molecular diffusivity of the scalar field like temperature, and $C_x \approx C/3$ is the streamwise Cox number. The Cox number C for temperature is the mean square temperature gradient over the square mean temperature gradient, and may be considered the ratio of turbulent to molecular diffusivity of the scalar.

The hydrodynamic phase diagram (HPD) concept was introduced by Gibson (1980) as a means of classifying microstructure patches according to their hydrodynamic state; that is, active, active-fossil, or fossil turbulence. A turbulence activity parameter $A_T = (\varepsilon/\varepsilon_o)^{1/2}$ for a microstructure patch with $\varepsilon_o \approx 13 D C_x N^2$ is plotted as a function of its Cox number $C = \overline{(\nabla T)^2}/(\nabla \overline{T})^2$ to determine the hydrodynamic state of the patch, where D is the molecular diffusivity and the stability frequency $N = \sqrt{(g/\rho) \partial\rho/\partial z}$. Most oceanic microstructure patches are fossilized by this measure, with $A_T < 1$, especially for large C. This method works because the viscous dissipation rates $\varepsilon$ for an isolated, fossilizing, turbulence patch decrease more rapidly with time than the temperature dissipation rates $\chi = 2DC(\nabla \overline{T})^2$. The areas under the velocity gradient spectra on the left are equal to $\varepsilon/6\nu$, and the areas under the temperature gradient spectra on the right are equal to $\chi/6D = C(\partial \overline{T}/\partial z)^2/3$. Thus, fossil vorticity turbulence persists much longer than fossil temperature turbulence, and velocity microstructure in fossilized patches appears negligible even though the temperature microstructure appears active. A similar hydrodynamic phase diagram was introduced by Imberger and Ivey (1991) based on extensive temperature and velocity microstructure measurements in lakes, reservoirs and the coastal ocean, clearly demonstrating the fossilization process.

Prandke et al. (1988) independently discovered the fossil turbulence phenomenon in the Baltic Sea using microconductivity probes and thermistors to detect small scale temperature, salinity and density, and a rugged and sensitive microshear sensor of their own invention at Wärnemünde in the former GDR. From vertically rising probes, they studied the diurnal formation and fossilization of surface mixed layers, and developed methods of estimating the ages of deeper microstructure patches from the relative decay of temperature and salinity microstructure, giving ages of the patches of ten hours or more for patches as small as a meter. Their patch ages were confirmed by temperature-salinity diagrams and velocity measurements they used to identify the geographical origins of the patches. At the 1995 workshop on "Fossil Turbulence and Hydropaleontology" at the IAPSO XXI General Assembly in Honolulu, HI, Adolph Stips and Hartmut Prandke showed individual patches plotted on hydrodynamic phase diagrams with the usual broad scatter of hydrodynamic states. Dissipation rates for patches detected in the Arkona Basin of the Baltic Sea were completely uncorrelated with the near bottom velocity, suggesting the patches of microstructure observed were fossils produced elsewhere, probably on the slopes of the basin. Dual probe techniques showed that correlations with cylindrical vertical anisotropy develop in the final stages of the fossil turbulence decay process, consistent with salt fingering mechanisms being active in Baltic Sea waters, which have strong contributions of both temperature and salinity to density. A full report is given in Stips, Prandke and Neumann (1998). At the same workshop Iossif Lozovatsky presented similar fossilization evidence from the Black Sea, where dissipation and overturn scale data from measured microstructure patches were displayed using hydrodynamic phase diagrams similar to Fig. 6.

The persistence time of fossil vorticity turbulence is estimated in Gibson (1980) as $\tau_u \sim N^{-1} (\varepsilon_o/\varepsilon_F)$ compared to the persistence time of fossil temperature turbulence of $\tau_T \sim N^{-1} (\varepsilon_o/\varepsilon_F)^{2/3}$. We see that the ratio $\tau_u/\tau_T \sim (\varepsilon_o/\varepsilon_F)^{1/3}$ increases with the event Reynolds number ratio $(\varepsilon_o/\varepsilon_F) = Re_O/Re_F$. Various estimates of microstructure patch ages are reviewed by Gregg (1987) who concludes that claims of such long persistence times of fossil turbulence "have been refuted" by Caldwell (1983), Dillon (1984), Elliott and Oakey (1976) and Gregg and Sanford (1980), leading to Misconception 9 of Table 1. This incorrect conclusion and its basis are discussed in Gibson (1987). Another suggested measure of oceanic microstructure age is the spectral Shannon entropy, Wijesekhara and Dillon (1997). Shannon entropy $S = -\sum p_i \ln p_i$ is a measure of the complexity of a random process of probability p, where i is the number of events. Thus constant p gives maximum $S = 1$. Wijesekhara and Dillon (1997) hypothesize that a normalized spectral measure $S_n = S/\ln M$, where $p_i = \phi_i/\sum \phi_i$ and M is the number of spectral bins, should increase with age as the cascade from large scales to small proceeds and the spectrum $\phi$ gets whiter (more constant over a wider range of wavenumbers). However, as shown in Fig. 4, spectra for both the velocity and temperature gradients of a microstructure patch as it fossilizes are initially wide and flat, so that $S_n$ is large, and become steeper and narrower, with smaller $S_n$, with increasing age, opposite to the Wijesekhara and Dillon (1997) prediction. Temperature gradient spectra measured in the powerful turbulence behind tidal sills in the Gulf of California, Gibson (1998, from Myrl Hendershott, personal communication), show $S_n$ values of 0.7, decreasing to 0.2 more than a hundred miles downstream several days later, with persistent vertical fossil-temperature-turbulence overturns up to 60 m and close agreement with the predicted shape and saturated spectral level for $\varepsilon_T$ of equation (7), Gibson (1980). In the initial stages of turbulence development $S_n$ might give a reliable qualitative measure of a growing patch's age, but $S_n$ will give no reliable measure or a false measure of either age or life expectancy of patches after fossilization begins.

As shown in Fig. 4, the trajectory of $A_T$ as the wake turbulence fossilizes is from the actively turbulent quadrant with $A_T > 1$, to the point of beginning fossilization at point 2 with $C = C_o$, toward the sloping line of complete fossilization with $\varepsilon = \varepsilon_F = 30\ \nu N^2$.

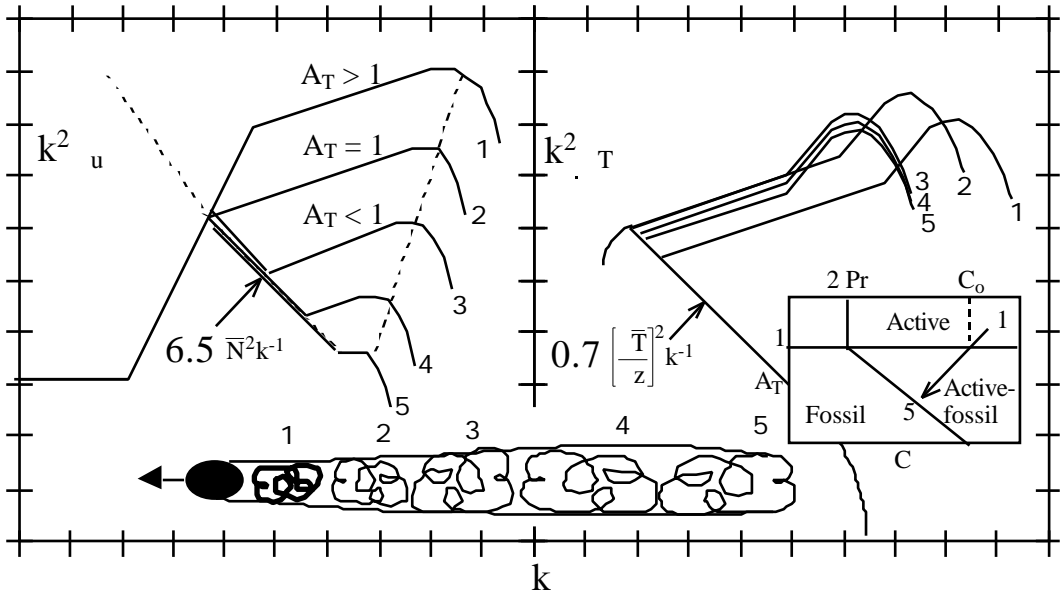

**Figure 4.** Dissipation spectra for velocity, left, and temperature, right, for active turbulence as it fossilizes, represented by the five stages of the stratified turbulent wake, bottom. Universal spectral forms for saturated internal waves from Gibson (1980) are shown. The trajectory on an $A_T$ versus C hydrodynamic phase diagram is shown in the right insert.

$A_T$ versus C HPDs have the advantage that the hydrodynamic state of a microstructure patch can be determined using only temperature measurements. However, a more complete representation of the stratified hydrodynamics of the turbulent event is provided by including the ambient stratification frequency N for the patch in the HPD, so that $A_T$ can be interpreted in terms of the Froude number of the patch normalized by $Fr_o$ at beginning of fossilization plotted versus the Reynolds number normalized by $Re_F$ of the patch at complete fossilization. Cox number C was replaced by the Reynolds number ratio $Re/Re_F = \varepsilon/\varepsilon_F$ in subsequent HPDs of Gibson (1986, 1996), using the Gibson (1980) dissipation rate at complete fossilization $\varepsilon_F = 30\ \nu N^2$, and $A_T$ was replaced by the Froude number ratio $Fr/Fr_o = (\varepsilon/\varepsilon_o)^{1/3} = A_T^{2/3}$. By either HPD, most oceanic microstructure patches fall in the active-fossil quadrant, with $Fr/Fr_o < 1$ and $Re/Re_F > 1$, after a short time period of active turbulence growth period from small to large scales with $Fr/Fr_O > 1$ and $Re/Re_F > 1$. Dominant actively turbulent patches in equatorial layers with $Re_O/Re_F > 10^6$ are inferred from HPD plots of 20-30 m fossilized patches, and $Re_O/Re_F > 10^4$ for abyssal layers from 10 m patches, but such rare patches have not been observed in their actively turbulent states, and are not expected to be detected at the present low sampling rates.

### 3.1 Definition of fossil turbulence

Many hydrophysical fields are scrambled by an actively turbulent event, and these "footprints" may persist independently for different periods, preserving identical or different subsets of the event information as fossil turbulence long after the turbulence, by any definition, has died away. Assuming the definition of active turbulence given previously, fossil turbulence may be defined as follows,

> **Definition:** *Fossil turbulence is defined as a fluctuation in any hydrophysical field produced by turbulence that persists after the fluid is no longer actively turbulent at the scale of the fluctuation.*

To discriminate between fossils preserved in different hydrophysical fields, the terminology "fossil vorticity turbulence", "fossil temperature turbulence", etc. is often convenient. This definition and the definition of turbulence given above are somewhat broader than those given by Gibson (1980) in order to include the crucially important phenomenon of horizontally polarized 2D turbulence and 2D fossil turbulence, Gibson (1990, 1991cd), which is probably the source of most 3D turbulent patches and their extremely intermittent lognormality. Considerations of intermittency and 2D turbulence are beyond the scope of the present paper.

### 3.2 Identification of fossil turbulence

Microstructure patches can easily be classified according to their hydrodynamic state using the hydrodynamic phase diagram shown schematically in Figure 5.

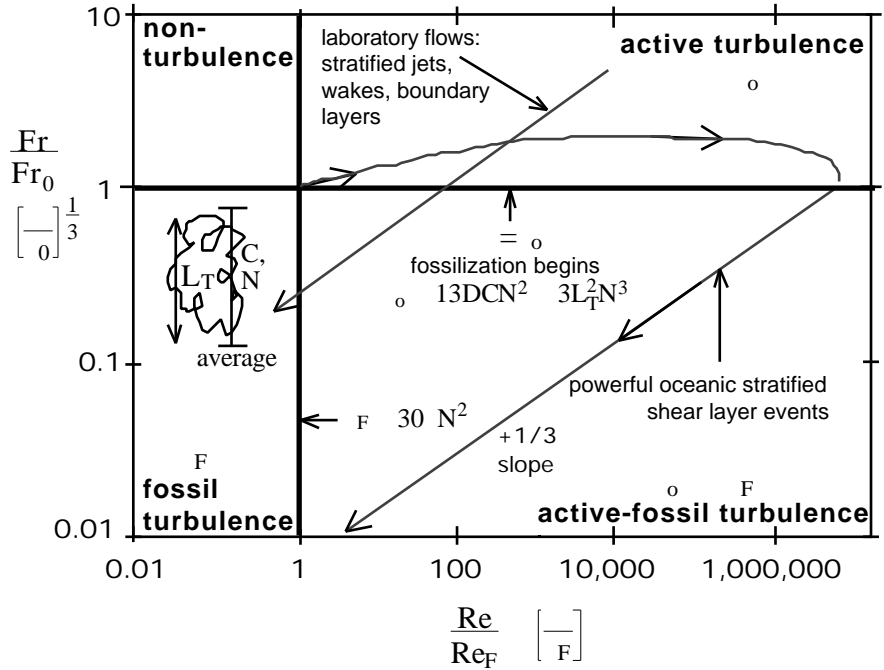

**Figure 5.** Hydrodynamic phase diagram used to classify microstructure patches according to their hydrodynamic state in a stratified fluid. The dissipation rate $\varepsilon$ is measured for the patch, and the normalized Froude number and Reynolds numbers computed from the expressions for $\varepsilon_o$ and $\varepsilon_F$ given on the diagram. Both the maximum Thorpe overturn scale $L_T$ and the Cox number C can be used to estimate $\varepsilon_o$. The ambient N for the patch can be used to find $\varepsilon_F$, where C and N must be averaged over vertical scales larger than the patch, as shown in the inserted sketch. Trajectories for the evolution of stratified laboratory and dominant oceanic turbulent events are indicated by arrows and dashed lines.

The ordinate in Fig. 5 is derived from the definition of Froude number $Fr = u/LN$, using $u \approx (\varepsilon L)^{1/3}$ for the characteristic velocity $u$ of turbulence from Kolmogorov's second universal similarity hypothesis. The ratio $Fr/Fr_o = (\varepsilon/\varepsilon_o)^{1/3}$ represents a comparison of the Froude number of the given patch with the Froude number of a completely turbulent patch having the same L and N at the point of fossilization, denoted by o-subscripts. The abscissa in Fig. 2 is derived from the definition of Reynolds number $Re = uL/\nu = c\varepsilon/\nu N^2$, using the same Kolmogorov expression for u and assuming the maximum turbulence scale in the patch is the Ozmidov scale $L_R = (\varepsilon/N^3)^{1/2}$, so that $Re_F = c30$ substituting $\varepsilon_F = 30\nu N^2$. The ratio $Re/Re_F$ is therefore just $\varepsilon/\varepsilon_o$ since the constant c cancels.

## 4. Fossil turbulence in the ocean

Turbulence in the ocean is extremely rare in space and time, and is rapidly damped by buoyancy and Coriolis forces. The kinetic energy of the turbulence and the variance of temperature, salinity, density, chemical species, and biological species produced by the turbulence stirring are not returned to their original positions but are gradually and irreversibly mixed later and elsewhere by the active-fossil and fossil turbulence motions produced when the active turbulence is damped, and by other processes such as double diffusion or shear of the fossil. Most mixing and most vertical and horizontal diffusion in the ocean are initiated by active turbulence, but the mixing and diffusion is completed by fossil turbulence processes and is spread to much larger volumes of fluid and to other ocean layers and regions as fossil turbulence within the patches, and by fossil turbulence internal waves and Coriolis-inertial waves radiated by the patches. Fossil turbulence processes are crucially involved in all aspects of mixing and diffusion in stratified and rotating fluids that become actively turbulent.

The most important of the misconceptions of Table 1 are the first two concerning the direction of the turbulence cascade and the definition of turbulence. Neither misconception provides any difficulties in non-stratified turbulence studies in quasi-equilibrium such as grid wakes, but they are crucial in oceanography since buoyancy forces dominate fossil vorticity turbulence (which is not turbulence by definition, from the right column of Table 1, but a unique class of internal waves). If turbulence cascades from large scales to small, as often assumed, then the signature of fossil turbulence is not unique since patches of microstructure with large vertical overturning scales and small dissipation rates might simply be large scale turbulence eddies that have not yet become turbulent.

The idea that turbulence cascades from large scales to small is a myth deeply ingrained in turbulence folklore. It is enshrined by the Richardson poem. It was mentioned by Kolmogorov (1941). The time required for large eddies to form the first small eddies is estimated by Lumley (1992). It is the basis of the Gregg (1987) stratified turbulence model where turbulence forms by the gravitational collapse of non-turbulent billows, which then vanish without any fossil turbulence trace in a time period of order $N^{-1}$. The Gregg (1987) model is based on an optical illusion; that is, laboratory Kelvin-Helmholtz billows in their initial stages, Thorpe (1987), appear to be inviscid but are not. When a thin density interface is tilted, boundary layers form and are likely to become turbulent before the billows appear under oceanic conditions, with maximum dissipation rate as the billows develop rather than zero as assumed by Gregg (1987). In fact, turbulence does not develop by gravitational collapse for billows under either laboratory or oceanic conditions, as assumed by Gregg (1987), but by shear instabilities before, during, and after the formation of the billows. Gravitational forces are irrelevant except to inhibit the initial billow formation, and later after the turbulent billows grow to the Ozmidov scale $L_{R_O}$ and fossilization begins. De Silva et al. (1996) show the billow breakdown by turbulence occurring before the billow reaches its maximum size, contrary to the Gregg (1987) model.

### 4.1 Summary HPD

Figure 6 shows a summary of oceanic microstructure classified according to hydrodynamic state of the microstructure patches using the normalized Froude versus Reynolds number HPD of Fig. 5.

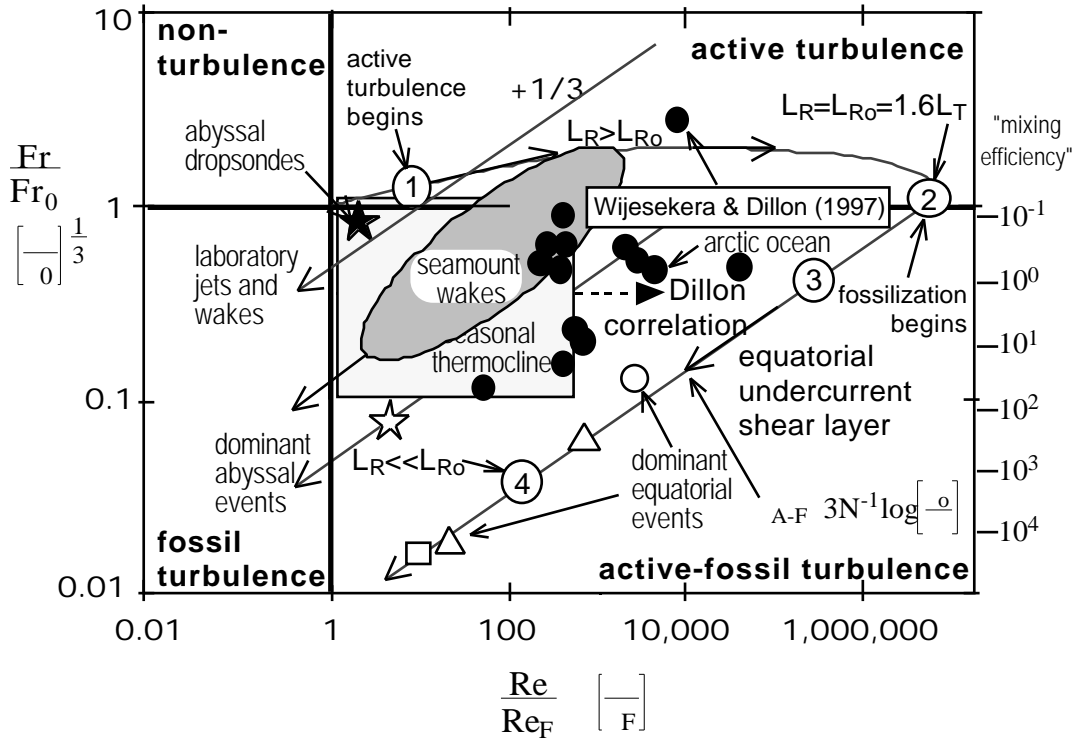

**Figure 6.** Hydrodynamic phase diagram for microstructure in the laboratory, seasonal thermocline, abyssal layers, seamount wakes, Arctic, and equatorial undercurrent (adapted from Gibson 1996). The Dillon (1984) correlation $L_R = L_{Trms}$ shown by the horizontal dashed arrow assumes seasonal thermocline patches are actively turbulent rather than active-fossil. Large equatorial patches with density overturns of twenty to thirty meters have been observed by Hebert et al. (1992) ◯, Wijesekera and Dillon (1991) △, and Peters et al. (1994) ☐, indicating $_o$ values in previous actively turbulent states more than $10^6$ $_F$ and $10^3$-$10^5$ larger than measured values of for these dominant equatorial events. Abyssal dropsonde profiles ★ Toole, Polzin and Schmitt (1994), are usually so sparse they fail to detect even the fossils of the dominant turbulence events, ☆ (inferred by Gibson 1982a). The "mixing efficiency" $(DCN^2/)$ = $_o/13$ shown on the right (Oakey 1982) can be $10^4$ larger than 100% for fossilized patches. Measurements of powerful patches in the Arctic, ● Wijesekera and Dillon (1997) confirm the fact that large, actively-turbulent microstructure events exist in the ocean interior, even though they are rare.

As shown in Fig. 6, the laboratory studies have much smaller $Re_o/Re_F$ values that most microstructure in ocean layers, corresponding to the weakest and smallest patches usually reported, such as in the seasonal thermocline indicated by the square box. The first most extensive measurements of microstructure were made during the MILE expedition, where simultaneous dropsonde and towed body measurements were possible. Since most microstructure patches found were relatively weak, Dillon (1984) proposed that possibly they were actively turbulent and that the universal constant had been incorrectly estimated in Gibson (1980). Dillon's correlation is shown by the horizontal arrow. When Dillon's correlation was proposed, few patches had been observed in their actively turbulent state except by towed bodies, despite laboratory confirmations of the Gibson (1980) universal constants by Stillinger et al. (1983) and Itsweire et al. (1986) in a stratified grid turbulence flow, as discussed by Gibson (1991d). However, measurements in the wake of Ampere Seamount, Gibson et al. (1994), show that oceanic microstructure indeed begins in the actively turbulent quadrant for patches sampled over the crest of the seamount, and decays into the active-fossil quadrant for patches sampled downstream of the crest, as indicated by the oval region and arrow. We see from Fig. 6 that ocean microstructure is not likely to be found in its original actively turbulent state, but in some stage of fossilization. Wijesekera and Dillon (1997) report only one fully active patch from their Arctic Sea data (dark circles).

## 4.2 Dark mixing paradox

Tom Dillon has suggested that the most important outstanding problem of physical oceanography today is the resolution of the "dark mixing paradox"; that is, unobserved mixing that must exist somewhere in the ocean to explain the fact that the ocean is mixed, by analogy to the dark matter problem of astrophysics, which is unobserved matter that must exist to overcome the centrifugal forces of rotating galaxies by gravity. The largest discrepancies between flux-dissipation rate estimates of vertical diffusivities and those inferred from bulk flow models are in strongly stratified layers of the upper ocean such as the seasonal thermocline and the equatorial undercurrent, and deep in the main thermocline, Gibson (1990, 1991a, 1996).

The discrepancies disappear when the evidence of vast undersampling provided by fossil turbulence is taken into account, and the extreme lognormal intermittency of the dissipation rates are used to estimate mean values, rather than using typical values that are closer to the mode rather than the mean. The mean to mode ratio of a lognormal random variable is $\exp(3\sigma^2/2)$, where $\sigma^2$ is the intermittency factor, or variance of the natural logarithm about the mean, of the probability distribution function. Baker and Gibson (1987) have shown that $\sigma^2$ for $\epsilon$ and $\chi$ in the ocean range from 3-7 in the ocean, showing that undersampling errors are probably in the range 90 to 36,000 if the fossil turbulence evidence is ignored and no attempt is made to compensate for intermittency. Gibson (1991a) shows that when the observed lognormality of dissipation rates in the deep thermocline is taken into account, there is no discrepancy between the vertical diffusivity of temperature inferred using the Munk (1966) abyssal recipe, and that inferred from the Gregg (1977) deep Cox number measurements.

## 4.3 Fossil turbulence pattern recognition by phytoplankton

It has been known for more than 60 years by marine biologists that phytoplankton growth is sensitive to turbulence, particularly the growth of dinoflagellates that cause red tides. Allen (1938, 1942, 1943, 1946ab) observed that red tides in La Jolla bay were preceded by at least a two week period of sunny days and light winds, and would not occur otherwise. Margalev (1978) suggested a "mandala" (a schematic representation of the cosmos in eastern art and religion) for phytoplankton survival where dinoflagellates are favored when turbulence levels are low and diatoms are favored when turbulence levels are high. Efforts to grow dinoflagellate species in the laboratory typically fail when the cultures are shaken vigorously for long periods, or aerated.

Using the Couette flow between two concentric cylinders to achieve a uniform rate-of-strain for phytoplankton cultures compared to a motionless control, Thomas and Gibson (1990ab, 1992) quantified the growth response of the red tide dinoflagellate Gonyaulax *polyedra* Stein. It was found that the viscous dissipation rate $\varepsilon^2$ for growth inhibition $\varepsilon_{GI}$ was about 0.2-0.3 cm$^2$ s$^{-3}$. Other dinoflagellate species tested also showed a negative response to continuous strain rates. In the case of *Prorocentrum micans*, negative growth rate was observed only for high light levels of 332-451 μE m$^{-2}$ s$^{-1}$ and not for lower values of 162-209 μE m$^{-2}$ s$^{-1}$, Tynan, Thomas and Gibson (in progress).

Gibson and Thomas (1995) report the effects of turbulence intermittency on dinoflagellate growth, based on laboratory studies and the field tests of Tynan (1993). The remarkable result was found for Gonyaulax *polyedra* that continuous strain rates with $\varepsilon > \varepsilon_{GI}$ have the same or greater negative effects on growth when applied for short time periods T greater than a minimum $T_{GI}$ during a day for several days, where $T_{GI}$ was in the range of only 5-15 minutes. Thus, intermittency of turbulence reduces the daily average $\varepsilon$ threshold values for growth inhibition by factors of 100 or more. Vigorous stresses applied for short periods (<< 5 minutes) were ignored by the dinoflagellate cultures, for example when the total culture was mixed together at least daily to determine an accurate average cell concentration during the ten day tests. Dissipation $\varepsilon$ values to cause maximum bioluminescence are about $10^5$ $\varepsilon_{GI}$ for Gonyaulax *polyedra* and are non-fatal, based on viscous stress levels of 10 dynes/cm$^2$, Rohr et al. (1997). Tynan (1993) found a negative correlation of the dinoflagellate population, after a three day lag, with significant wave height and wave velocity, and the opposite response for the diatom population. Neither dinoflagellate nor diatom populations were strongly correlated with wind speed. Thomas, Tynan and Gibson (1997) report positive effects of laboratory turbulence on diatom growth, although the effects of intermittency on diatoms have not been tested in the laboratory. How can these patterns of phytoplankton growth response be explained?

Gibson and Thomas (1995) propose the possibility that phytoplankton have evolved the ability to recognize patterns of turbulence-fossil-turbulence in the surface layer of the sea and adjust their growth rates in response to these patterns to optimize their competitive advantages with respect to swimming ability. Dinoflagellates can swim but diatoms cannot. Diatoms require turbulence in the surface layer to bring them up to the light. When the surface layer is strongly stratified by sunny days and long periods without turbulence, dinoflagellates can swim toward the light and bloom if adequate nutrients are available, but diatoms will sink out. Breaking surface waves produce intermittent turbulence patches in the euphotic zone that have very different $\varepsilon$ values as a function of time depending on whether the fluid is stratified or non-stratified. If the surface layer is not stratified the dissipation spectrum k$^2$ $\phi_u$ is the universal Kolmogorov form shown in Fig. 1, curve 1, with +1/3 slope and persistence time of only a few seconds corresponding to a few overturn times of the largest eddies. If the surface layer is stratified, however, the dissipation rate is preserved above ambient by the fossilization process. From Gibson (1980) the decay time T for fossil vorticity turbulence is about N$^{-1}$($\varepsilon_o/\varepsilon_F$) = N$^{-1}$(Re$_o$/Re$_F$). Breaking surface waves have very large $\varepsilon$ values compared to those measured in the ocean interior which are generally much less than $\varepsilon_{GI}$ values indicated for any dinoflagellate species. Dissipation rates measured in the surf zone were found to be $\varepsilon$ = 0.5 to 500 cm$^2$ s$^{-3}$ by George et al. (1994), which could give $\varepsilon$ > $\varepsilon_{GI}$ for periods T much longer than T$_{GI}$ if the surface layer is strongly stratified so that $\varepsilon_F$ > $\varepsilon_{GI}$. However, the appearance of breaking surface waves in one location more than once a day for several days is a good indication that a sea change is in progress from a strongly stratified surface layer, favorable to dinoflagellates that can swim, to a turbulent mixed surface layer, favorable to diatoms that cannot. The model is illustrated by an example on a hydrodynamic phase diagram in Figure 7 using threshold values $\varepsilon_{GI}$ and T$_{GI}$ measured for dinoflagellate *Gonyaulax polyedra* Stein.

In Fig. 7 it is assumed that $\varepsilon_{GI}$ = 0.3 cm$^2$ s$^{-3}$, so that N ≈ 1 rad/s to give $\varepsilon_F$ = 0.3 cm$^2$ s$^{-3}$. This is very strong stratification for the ocean, but could happen near the surface under red tide conditions. For a breaking wave patch to persist for 15 minutes (900 s) with this stratification requires ($\varepsilon_o/\varepsilon_F$) ≈ 1000, so $\varepsilon_o$ ≈ 300 cm$^2$ s$^{-3}$ or greater, less than values measured by George et al. (1994) in the surf zone. A shaded zone of possible phytoplankton growth effects is shown in Fig. 7. To the left of the zone, surface wave patches will not persist long enough to have an effect. To the right of the zone, the surface wave breaking might be so powerful that the turbulent jet would penetrate below the euphotic layer. Larger N values require larger $\varepsilon_o$ values, and both are physically unlikely. Smaller N values might not give patches that persist long enough with $\varepsilon$ > $\varepsilon_{GI}$, although the question of how dissipation rates evolve with time in a fossil turbulence patch is an important unsolved problem. A semi-empirical estimate of the time $\tau_{A-F}$ for $\varepsilon_o$ to decay to a value $\varepsilon$ in the active-fossil quadrant is given by the expression in Fig. 5.

The active-fossil-turbulence pattern recognition proposal of Gibson and Thomas (1995) is intended as a working hypothesis that explains the presently available data for the response of phytoplankton to turbulence and turbulence intermittency. Other scenarios may exist, but have so far not been put forth.

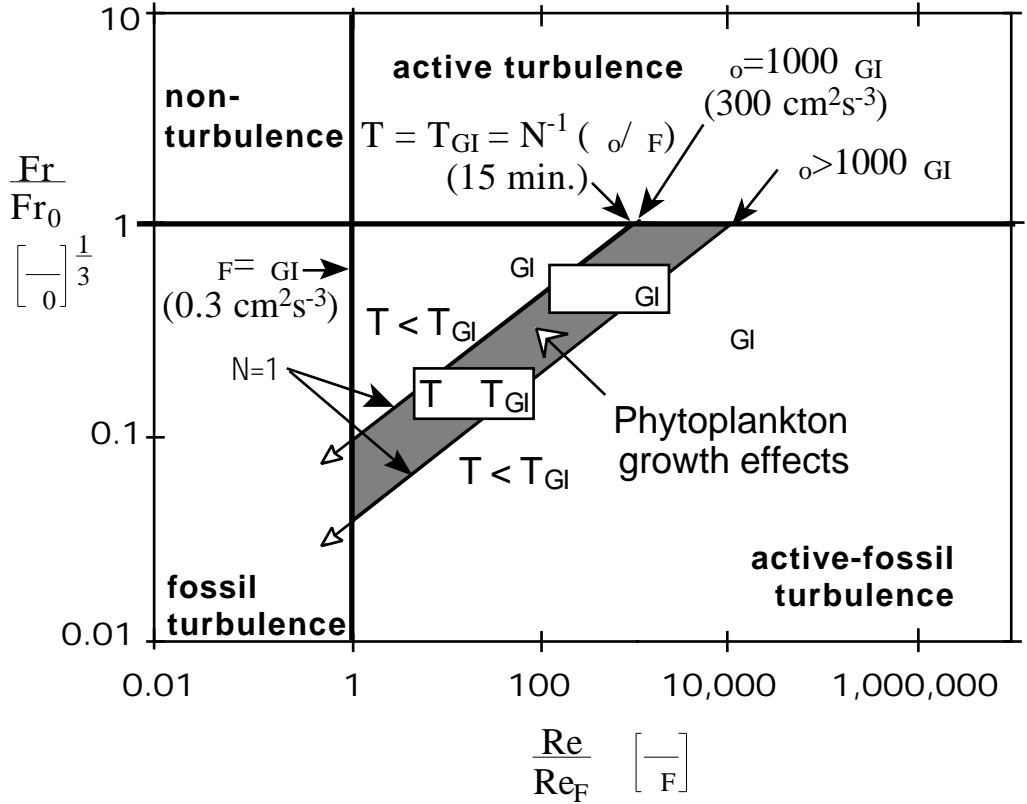

**Figure 7.** Hydrodynamic phase diagram for phytoplankton growth effects based on the *G. polyedra* laboratory results. The constraints $\varepsilon > \varepsilon_{GI}$ for time $T > T_{GI}$ are satisfied only for $N = 1$ rad/s and turbulence at fossilization $\varepsilon_o$ values $\geq 1000 \varepsilon_{GI}$.

## 5. Fossil turbulence in cosmology

As suggested by Gamov (1954), the most ancient fossils of turbulence are likely to be cosmological, assuming turbulent fluid mechanical processes in the early epochs of the universe had an influence on cosmological structure formation that persists to the present.

### 5.1 Scales of self-gravitational condensation

In a review of turbulence in natural fluids, Gibson (1996) suggests the Jeans (1902, 1929) theory of gravitational instability is unreliable as a means of predicting the mass and length scales of self-gravitational condensation after the big bang, both in cosmology and in galactic star formation. Jeans used a linear perturbation stability analysis of the mass and momentum conservation equations to predict the minimum scale $L_J$ of condensation in a uniform gas of density $\rho$ should be

$$L_J = V_S/(\rho G)^{1/2}; \quad (RT/\rho G)^{1/2}; \quad (p/\rho^2 G)^{1/2} \tag{9}$$

where $V_S$ is the speed of sound in the gas, G is Newton's constant of gravitation, R is the gas constant, and p is the pressure. Linear theory is inappropriate for self-gravitational condensation because it neglects the possibility of turbulence, turbulent mixing, and especially non-acoustic density maxima and minima produced by turbulent mixing, Gibson (1968ab), which are absolutely unstable to condensation and void formation at the largest scale permitted by viscous or turbulent forces or by molecular diffusivity. The Jeans theory may be relevant to self-gravitational condensation, but only briefly during the initial phases of structure formation after the hot plasma produced by the Big Bang at time t = 0 cooled to 3000 K from the expansion of the universe to form neutral gas at t = 300,000 years. The density of the "primordial gas" formed was $10^{-18}$ kg m$^{-3}$ at this time from Einstein's general relativity theory, Weinberg (1972, Table 15.4).

The first approximation of $L_J$ on the right of (9) represents the scale $L_{IC}$ of the initial condensation of the primordial gas while the linear approximations of Jeans theory are valid, and depends only on the composition of the gas, to give R, the density , and the temperature T. The primordial gas consists of 25% $^4$He by mass, so R 4740 m$^2$ s$^{-2}$ K$^{-1}$. Thus, $L_{IC}$ of the primordial gas is 4.6x10$^{17}$ m substituting these values and G = 6.72x10$^{-11}$ m$^3$ s$^{-2}$ kg$^{-1}$, giving an initial condensation mass $M_{IC}$ $L_{IC}^3$ = 9.8x10$^{34}$ kg. This is nearly 10$^5$ solar masses, and is often interpreted as the explanation for spherically symmetric globular clusters of stars in galaxy halos which usually contain about this mass, nearly a million small ancient stars, and densities near the primordial gas value. Such an explanation is an example of hydropaleontology, since the interpretation implies the hydrodynamic state of the primordial gas was sufficiently quiet that turbulence did not prevent condensation at the turbulent Schwarz scale $L_{ST}$ = $^{1/2}$/( G)$^{3/4}$ where turbulent inertial vortex forces ( L)$^{2/3}$L$^2$ match self gravitational forces $^2$GL$^4$, Gibson (1996). This puts an upper bound on for the primordial gas of 1.6x10$^{-7}$ m$^2$ s$^{-3}$, comparable to dissipation rates in the ocean interior.

The second approximation of $L_J$ in (9) is simply a hydrostatic scale $L_{HS}$ obtained by equating the interior pressure force pL$^2$ to the gravitational force $^2$GL$^4$ of a gravitationally bound fluid object in a perfect vacuum. Numerous fallacious derivations of the Jeans scale assert that condensations on scales smaller than $L_J$ are prevented by "pressure support" or "thermal support" using this expression; for example, Weinberg (1977), Shu (1982), Silk (1989). Hydrostatic pressures do not inhibit gravitational condensation on non-acoustic density nuclei at scales smaller than $L_J$ if turbulent, viscous, or magnetic forces do not prevent it and if the molecular diffusivity of the density is small enough. Balancing these forces with gravitational forces, and the diffusion velocity D/L with the gravitational velocity L( G)$^{1/2}$ of density gives the Schwarz gravitational condensation scales $L_{SX}$,

$$L_{ST} = {}^{1/2}/( G)^{3/4}, L_{SV} = ( / G)^{1/2}, L_{SM} = [(H^2/ )/ G]^{1/2}, L_{SD} = (D^2/ G)^{1/4} \quad (10)$$

termed turbulent, viscous, magnetic and diffusive Schwarz scales, where the magnetic pressure H$^2$ is the local variance of the magnetic field **H**. These scales $L_{SX}$ may be larger or smaller than the Jeans scale $L_J$, and condensation occurs at the largest $L_{SX}$ independent of the value of $L_J$. $L_{ST}$ is closely analogous dynamically to the Ozmidov scale $L_R$. Application of this new self-gravitational condensation theory by Gibson (1996) leads to the conclusion that the earliest fossils of very weak turbulence are the masses of proto-superclusters, proto-clusters, and proto-galaxies formed in the plasma epoch between 30,000 and 300,000 years after the big bang. These masses preserve information about the primordial values of , D, and in this non-oceanographic application of hydropaleontology using evidence from fossil turbulence and fossil non-turbulence.

## 5.2 Dark matter

According to standard cosmological models, Silk (1994), only one part in a thousand of the matter in the universe is luminous as stars and quasars. The rest is dark matter. Dark matter can be baryonic (ordinary) or non-baryonic (with small collision cross-section, like neutrinos but with a rest mass). Only 3% can be baryonic from the observed chemical composition and nucleosynthesis models. Because the diffusivity and viscosity coefficients of non-baryonic matter are so much larger than those of baryonic matter its $L_{SV}$ and $L_{SD}$ scales are enormous, so that the non-baryonic dark matter can only condense to form halos at supercluster scales and not on galaxies, Gibson (1996, 1997). By this theory, dark matter in galaxies is dominated by baryonic condensations at the plasma to gas transition, at 300,000 years ($10^{13}$ s).

The weighted average viscosity µ of the primordial gas mixture of hydrogen and helium extrapolated to 3000 K with the expression $\mu = \mu_{293}(T/293)^{0.68}$ is $5.6 \times 10^{-5}$ kg m$^{-1}$ s$^{-1}$ for the primordial gas. The density of the condensing primordial gas must have been larger than the universe average $= 10^{-18}$ kg m$^{-3}$ because of condensations at larger scales, possibly about $10^{-17}$ kg m$^{-3}$ from the observed density of globular clusters. The minimum rate-of-strain for the expanding universe was $= 1/t = 10^{-13}$ s$^{-1}$. The primordial condensation mass at the viscous Schwarz scale $L_{SV}$ is

$$M_{SV} = \ L_{SV}^3 = (\ \mu\ /G)^{3/2}/\ ^2 \qquad (10)$$

giving $M_{SV}$ in the range $2.4 \times 10^{23}$ kg for $= 10^{-17}$ kg m$^{-3}$ to a maximum value of $2.4 \times 10^{25}$ kg for $= 10^{-18}$ kg m$^{-3}$, about ten orders of magnitude smaller than the Jeans mass $M_J$. Rather than a mass of millions of stars, the first condensation mass is that of the moon or the earth.

These small condensates of the primordial gas are termed "primordial fog particles" (PFPs), Gibson (1996), and constitute the original material of construction for all subsequent structures, such as stars and the gassy planets of the outer solar system. Since most PFPs have not yet aggregated to form stars or anything else, they persist as 15 billion year old fossil turbulent remnants of the plasma-gas transition, and comprise most of the baryonic dark matter of galaxies along with the ashes of burned out stars such as black holes, interstellar dust, and humans. About $10^{19}$ PFPs exist in the Milky Way Galaxy along with $10^{11}$ stars. They are separated by $10^{14}$ m, and are cold and dark. Evidence of their existence can be found in the vicinity of exploding stars using the high resolution of the Hubble Space Telescope; for example, the 3500 "cometary globules" imaged in the Helix planetary nebula by O'Dell and Handron (1995), and the 2000 "knots" of the T Pixidis recurrent Nova observed by Shara et al. (1997). Mass estimates for the objects are in the PFP range. From microlensing of the twin images of the first discovered gravitationally lensed quasar TwQSO, Schild (1996) concludes that the mass of the lensing galaxy is dominated by numerous "rogue planets" of mass about $10^{25}$ kg, consistent with the Gibson (1996) interpretation that most galactic mass consists of non-aggregated PFPs. Catalano (1997) emphasizes the need for a theoretical explanation of these observed "rogue planets". As the British astronomer Arthur S. Eddington said, "Never trust an observation until it is confirmed by theory".

## 6. Summary and conclusions

Most of the observed microstructure patches that contribute significantly to the vertical diffusion in sampled layers of the ocean are found to be fossilized at the largest scales from their classification as active, active-fossil, and fossil turbulence using either $A_T$ versus $C$ or $Fr/Fr_o$ versus $Re/Re_o$ hydrodynamic phase diagrams, both HPDs based on the fossil turbulence theory of Gibson (1980), as shown in Figures 4 and 6. Consequently, vertical diffusivities for these layers estimated without taking fossil turbulence effects into account are unreliable. Failure to recognize fossil turbulence by the oceanographic microstructure community is attributed in Table 1 to a variety of common misconceptions and myths about turbulence. The most egregious and venerable of these is the unfortunate concept that turbulence cascades from large scales to small. Such turbulence behavior is widely accepted but never happens, despite the Kelvin-Helmholtz billow and wingtip vortex phenomena that might give this illusion. This generally harmless classical myth of quasi-stationary laboratory turbulence renders the signature of stratified fossil turbulence ambiguous because the signature of fossil turbulence; that is, a microstructure patch with large vertical overturns in a stratified fluid but with dissipation too small to produce the overturns ($\varepsilon << \varepsilon_o \approx 3 L_T^2 N^3 \approx 13 DCN^2$) could be duplicated by these mythical inviscid large eddies.

As shown in Fig. 6, the fossil turbulence signature is observed in most ocean microstructure patches, especially the largest, dominant, patches of a layer. As mentioned, the signature would not be unique to fossil turbulence if large turbulent eddies ever formed first and small ones formed later, reversing the turbulence cascade direction seen in jets, wakes, and boundary layers where small eddies obviously form first and large ones later. Even if the mythical large-to-small turbulence cascade direction were possible, it would be an amazing coincidence if nearly all turbulence patches in the ocean were sampled only in their initial stages before small scales had time to develop, since this stage clearly occurs only during a tiny fraction of the total time such turbulence events and their fossils exist; for example, in the cases of Kelvin-Helmholtz billows and wingtip vortices.

The large-to-small turbulence cascade is untenable by the present definition of turbulence, Gregg (1987) and Lumley (1992) to the contrary notwithstanding, although this model might seem plausible for the many turbulent flows that extract their energy from irrotational flows at larger scales by a non-turbulent cascade such as that shown in Fig. 2. In Section 2 the theory of turbulence is discussed from first principles, the shear instability of turbulence due to inertial-vortex forces is described and shown in Fig. 1, and a precise definition of turbulence is given which makes it clear that this irrotational cascade of energy from large scales to small is non-turbulent because the vorticity $\omega$ is zero so that the inertial-vortex forces $\mathbf{v} \times \omega$ of the flow are zero.

In Section 3, fossil turbulence theory is reviewed, fossil turbulence is defined, universal spectral forms derived in Gibson (1980) are given in Fig. 4, and the use of hydrodynamic phase diagrams to classify stratified fossil turbulence for laboratory and oceanic studies is summarized. Several huge (20-30 m) patches from the strongly stratified shear layer above the equatorial undercurrent show measured $\varepsilon$ values are as much as $3 \times 10^5$ times smaller than $\varepsilon_o$ values for the patches. Taken together they show the dominant turbulence events in this highly sampled but vastly undersampled layer have $\varepsilon_o/\varepsilon_F$ values greater than $10^6$, consistent with the largest

Gibson, C. H., Fossil turbulence revisited, Journal of Marine Systems, 21(1-4), 147-167, 1999intermittency factors $^2$ 7 in the world's oceans reported by Baker and Gibson (1987) and confirming the large dissipation rates reported in this layer from towed body turbulence measurements of Belyaev et al. (1974), Williams and Gibson (1974), Artemyeva et al. (1989) and Lilover et al. (1993).

Fossil turbulence theory have been applied in this paper to the dark mixing paradox of ocean mixing, the growth response of marine phytoplankton to turbulence and turbulent intermittency, and the dark matter paradox of cosmology, as examples of the wide range of scientific problems where hydropaleontology may be useful.

**References**
Allen, W. E. 1938. "Red water" along the west coast of the United States in 1938. Science, 88: 55-56.
Allen, W. E. 1942. Occurrences of "red water" near San Diego. Science, 96: 471.
Allen, W. E. 1943. Red water in La Jolla Bay in 1942. Trans. Amer. Microscop. Soc., 62: 262-264.
Allen, W. E. 1946a. "Red Water" in La Jolla Bay in 1945. Trans. Amer Microscop Soc., 65: 149-153.
Allen, W. E. 1946b. Significance of "red water" in the sea. Turtox News, Vol. 24, No. 2.
Artemyeva, T. S., I. D. Lozovatsky & V. N. Nabatov, Turbulence generation in the core of the Lomonosov Current in the vicinity of local frontal zones, Oceanology, 29, 41-48, 1989.
Baker, M. A. and C. H. Gibson 1987. Sampling turbulence in the stratified ocean: Statistical consequences of strong intermittency, J. of Phys. Oceanogr., 17, 10, 1817-1837.
Belyaev, V. S., M. M. Lubimtzev and R. V. Ozmidov 1974. The rate of dissipation of turbulent energy in the upper layer of the ocean, J. Phys. Oceanogr., 5:499-505.
Caldwell, D. R., T.M. Dillon, J.M. Brubaker, P.A. Newberger, and C.A. Paulson 1980. The scaling of vertical temperature gradient spectra, J. Geophys. Res., 85, C4, 1917-1924.
Caldwell, D. R. 1983. Oceanic turbulence: Big bangs or continuous creation?, J. Geophys. Res., 88(C12), 7543-7550.
Catalano, P. 1997. On the trail of Rogue Planets, Astronomy, Vol. 25, No. 12, 36-41.
De Silva, I. P. D., H. J. S. Fernando, F. Eaton, and D. Hebert 1996. Evolution of Kelvin-Helmholtz billows in nature and laboratory, Earth and Planetary Science Letters, 143, 217-231.
Dillon, T. R. 1984. The energetics of overturning structures: Implications for the theory of fossil-turbulence, J. Phys. Oceanogr., 14, 541-549.
Elliott, J. A. and N. S. Oakey 1976. Spectrum of small-scale oceanic temperature gradients, J. Fish. Res. Board Can., 33, 2296-2306.
Frisch, U., *Turbulence: The Legacy of A. N. Kolmogorov*, Cambridge University Press, UK, 1995.
Gamov, G. 1954. On the formation of protogalaxies in the turbulent primordial gas, Proc. Nat. Acad. Sci., 40, 480-484.
Gargett, A.E., T.B. Sanford, and T.R. Osborn 1979. Surface mixing layers in the Sargasso Sea. J. Phys. Oceanogr., 9: 1090-1111.
25